\documentclass[useAMS,usenatbib]{mnras}
\usepackage{multicol}
\usepackage{multirow}
\usepackage{natbib}
\usepackage{amsmath}
\usepackage{graphicx}
\usepackage{newtxtext,newtxmath}
\usepackage{xcolor}
\usepackage{pdflscape}
\usepackage{xspace}

\title[Pre-supernova feedback in spherical clouds]{Simulations of pre-supernova feedback in spherical clouds}
\author[Kourniotis et al.]{
M. Kourniotis$^{1}$\thanks{E-mail: kourniotis@asu.cas.cz},
R. W\" unsch$^{1}$,
S. Mart\'{i}nez-Gonz\'{a}lez$^{2}$,
J. Palou\v{s}$^1$,
G. Tenorio-Tagle$^2$,
and S. Ehlerov\'{a}$^1$\\
$^{1}$Astronomical Institute, Czech Academy of Sciences, Bocni II 1401, 141 31 Prague, Czech Republic\\
$^{2}$Instituto Nacional de Astrof\'isica, \'Optica y Electr\'onica, AP 51, 72000 Puebla, M\'exico}

\begin{document}

\newcommand{\ut}[1]{_{\text{#1}}}
\renewcommand{\sun}[1]{#1_{\odot}}
\newcommand{\MSun}{M$_\odot$\xspace}
\newcommand{\avir}{\alpha\ut{vir}}

\date{Accepted XXX. Received YYY; in original form ZZZ}
 
\pagerange{\pageref{firstpage}--\pageref{lastpage}}
\pubyear{2022}

\maketitle

\label{firstpage}

\begin{abstract}
We present a one-dimensional radiation-hydrodynamic model of a spherically symmetric cloud evolving under the influence of the self-gravity and the feedback from a star cluster forming in its centre. On one hand, the model is simple due to its 1D geometry, on the other hand, the feedback includes the ionising radiation, stellar winds and the radiation pressure acting on gas and dust. The star cluster is formed from the gas flowing into the cloud centre and the feedback parameters are determined from stellar evolution models and the cluster star forming history. The model is compared to the semi-analytic code \textsc{warpfield} implementing similar physical processes and exploring the scenario that the young cluster R136 in the Large Magellanic Cloud was formed due to re-collapse of the shell formed by the previous generation star cluster. A good qualitative agreement is found, however, $3 - 4$ times higher stellar mass is needed to disrupt the cloud in our model, because it takes into account (contrary to \textsc{warpfield}) self-gravity of the cloud surrounding the shell. We use the model to explore star formation in clouds with different mass, radius and density profile measuring their star formation efficiency (SFE), i.e. the fraction of the cloud mass converted to stars. We found that SFE is a function of a single parameter, $\mathrm{log(SFE)} \propto -n\ut{hm}^{-0.46}$, with $n\ut{hm}$ being the cloud mean particle density within its half-mass radius. Furthermore, we found that the feedback efficiency, i.e. a fraction of the feedback energy retained by gas, has a nearly constant value $\sim 10^{-3}$.

\end{abstract}

\begin{keywords}
hydrodynamics -- stars: formation -- stars: winds, outflows -- ISM: clouds -- \ion{H}{II} regions
\end{keywords}

\section{Introduction}

The cold gas assembled in the giant molecular clouds (GMC) serves as the building block of stars and star clusters. The efficiency with which the gas converts into stars, however, is long known to be limited based on the large depletion times that well exceed the dynamical ones \citep{1979ApJ...229..578S,2008AJ....136.2846B,2009ApJS..181..321E,2018ApJ...863L..21K}. When normalized to the free-fall time of the cloud, the star formation efficiency (SFE) has been observed to be as low as $0.3\%$ \citep{2018ApJ...861L..18U,2017ApJ...846...71L}, with theoretical predictions showing it can reach up to $10\%$ \citep{2016ApJ...826..200S,2021ApJ...911..128K}. The observed SFEs essentially reflect the dynamical conditions that regulate the global properties of GMCs in a diversity of environments. On the other hand, the simulated SFEs are largely dependent on the inclusion and the prescription of mechanisms that not only support clouds from gravitational collapse, but can effectively evacuate the sites of star formation.

Theoretical studies that rely solely on the dynamical state of the clouds (with the inclusion of magnetic fields) for regulating star formation, report values of SFE per free-fall time that are well above the observed ones \citep{2011ApJ...730...40P,2015ApJ...800...49L}. The magneto-hydrodynamic simulations of turbulent clouds by \cite{2019MNRAS.488.1501G} predicted an almost 100$\%$ SFE, which however, was substantially reduced to only a few percent when stellar feedback was included. In the same vein, \cite{2011MNRAS.417..950H} applied a stellar feedback model in high-resolution disc galaxy simulations and they were able to reproduce the observed Kennicutt–Schmidt relation over a wide range of surface densities $-$the exclusion of feedback resulted in excessive, by orders of magnitude, star formation rates. \cite{2013MNRAS.432..653D} suggested that cloud dispersion is achieved by the synergistic action of feedback and dynamics of the environment, each contributing differently at the different scales and galactic radii. This combination is also believed to regulate the cloud morphologies, velocity dispersions, and virial parameters, such that the latter do not drive the star formation in GMCs but rather, they are dependent on it \citep{2011MNRAS.413.2935D,2012MNRAS.420.3264G,2013MNRAS.435.1701C}.

The stellar feedback appears to be the key factor for regulating star formation in GMCs by clearing the leftover gas from newborn star clusters. This is well demonstrated by the spatial small-scale (<100 pc) decorrelation between tracers of molecular gas and of star formation \citep{2010ApJ...722L.127O,2018ApJ...863L..21K,2019Natur.569..519K,2021ApJ...918...13S}. The duration of the phase during which the clusters are embedded within the parent cloud spans only a few Myr \citep[e.g.][]{2015MNRAS.449.1106H,2017A&A...601A.146C,2019MNRAS.490.4648H}, suggesting clearance of the gas by mechanisms preceding the first supernova (SN) explosions \citep{2021ApJ...911..128K,2020MNRAS.493.2872C,2022MNRAS.509..272C}. Ultimately, most clusters are observed to dissociate from their parent clouds in less than 10 Myr \citep{2018MNRAS.481.1016G,2019MNRAS.483.4707G,2022MNRAS.516.4612T}.

The efficiency of the separate feedback forces, namely the photoionization, the ram pressure from the stellar winds, the direct or reprocessed radiation pressure acting on gas and dust, and the mechanical energy deposited by SNe, has been subject of investigation by both analytical studies and hydrodynamic simulations. Long considered to be important for the growth of \ion{H}{II} regions is the photoevaporation driven by the thermal pressure of the ionized gas \citep{1979A&A....71...59T,1979MNRAS.186...59W,1990ApJ...354..529B,2002ApJ...566..302M}. It drives the mass loss in clouds with low surface density \citep{2018ApJ...859...68K}, yet leaving bound massive clouds ($>10^{5}$ M$_{\odot}$) dynamically unaffected \citep{2012MNRAS.424..377D}. The compressed swept-up gas leads to the formation of dense structures that trigger star formation, though only for the short while \citep{2012MNRAS.427..625W,2020MNRAS.495.1672B}. The effect of radiation pressure was studied by \cite{2009ApJ...703.1352K}, who concluded that it is particularly important for the expansion of \ion{H}{II} regions surrounding high-mass clusters. More recently, it has been pointed that radiation pressure can dominate the thermal pressure of the warm gas only locally and soon after star formation \citep{2020MNRAS.498.4906B,2021MNRAS.501.4136A}. When compared to the ram pressure exerted by the shocked winds, the opportunity of radiation pressure to drive the shell dynamics is again confined at the early phases \citep{2016MNRAS.462.4532G}, and in cases when the ambient medium is rather dense \citep{2013ApJ...765...43S,2014ApJ...785..164M}. Observations also favor the dominance of photoionization and winds in shaping the morphology of \ion{H}{II} regions \citep[e.g.][]{2011ApJ...738...34P,2019MNRAS.486.5263M}. At later evolutionary phases, the contribution from SNe is believed to be less potent in clearing the star-forming gas, as most, if not all of the mechanical energy injected, is shown to escape into the wider environment through low-density channels, which have been carved by the early feedback forces \citep{2013MNRAS.431.1337R,2017MNRAS.464.3536R,2020MNRAS.493.4700L}. Nevertheless, SNe are believed to indirectly impact star formation dynamics by establishing and maintaining large-scale turbulence in the cloud \citep[e.g.][]{2016ApJ...822...11P}.

When the gravitational potential of the cloud is strong enough, or when feedback diminishes and becomes unable to drive the shell expansion e.g. due to leakage of energy, successive star formation episodes are possible to occur following re-collapse of the gas in the centre \citep[e.g.][]{2019A&A...622A..48R}. Based on this scenario, \citet[][hereafter R18]{2018MNRAS.473L..11R} utilized their analytical one-dimensional code \textsc{warpfield}, which describes the expansion of \ion{H}{II} shells \citep{2017MNRAS.470.4453R}, in order to explore the parameters that allowed the formation of the young cluster R136 in the Large Magellanic Cloud (LMC). The bound cluster is located in the star-forming region of 30 Doradus and is embedded within the older and dispersed population NGC 2070, consistent with a scenario in which star formation propagated inwards \citep{1999A&A...347..532S,2011ApJ...739...27D}. The age separation between the two distinct populations is estimated to exceed 1 Myr \citep{1998ApJ...493..180M,1999A&A...347..532S,2012ApJ...754L..37S}. This offset in time allowed R18 to propose a low threshold for the cloud density, $\sim$500 cm$^{-3}$, that could have led to re-collapse of the gas due to inefficient feedback from NGC 2070. The assumption by R18 that the natal cloud is in virial equilibrium, however, sets an important limitation to the model because it overrates the strength of the stellar feedback.

Thus stellar feedback acts to disrupt the host molecular cloud,  while also regulates its re-collapse. This dual role is the motivation for the current study. We present the results from one-dimensional radiation-hydrodynamic simulations for describing the interplay between the pre-SN feedback forces and the self-gravity of a cold gas cloud. By neglecting here any turbulent forces or external dynamics that would support the cloud against its collapse, we focus on the stellar feedback as the dominant mechanism to regulate the SFE. The objective of the study is twofold. First, we compare the predictions by R18 on the expansion of \ion{H}{II} shells with those for a self-gravitating cloud, and refine the parameters that may have enabled R136 to be formed upon re-collapse of the gas. Second,  we proceed to self-consistently infer the SFE, by allowing stars to be formed dynamically in the centre of the cloud until the feedback is able to clear the remaining gas. We explore density models that follow a Schuster profile to enable growth of the central cluster throughout different levels of the cloud compactness. The paper is structured as follows: in Section \ref{model}, we describe the physical model of the self-gravitating cloud, the individual feedback processes, and the adopted simulations setups. In Section \ref{code}, we introduce the numerical code that implements the physical processes. We present the results of the different setups in Section \ref{results}, and draw a link between the gas density and SFE. The limitations following our presented model are discussed in Section \ref{discuss}. The conclusions of the study are given in Section \ref{concl}.

\section{Physical model}
\label{model}

The physical model used in this work consists of a spherically symmetric cloud and a star cluster in its centre providing feedback in the form of radiation and stellar winds. The star cluster can either be inserted into the model initially (setup~A), or it can be formed during the calculation from the gas flowing into the central region (setup~B).

\subsection{Gaseous cloud}
\label{cloud}

\begin{figure}
\centering
\includegraphics[width=8.5cm]{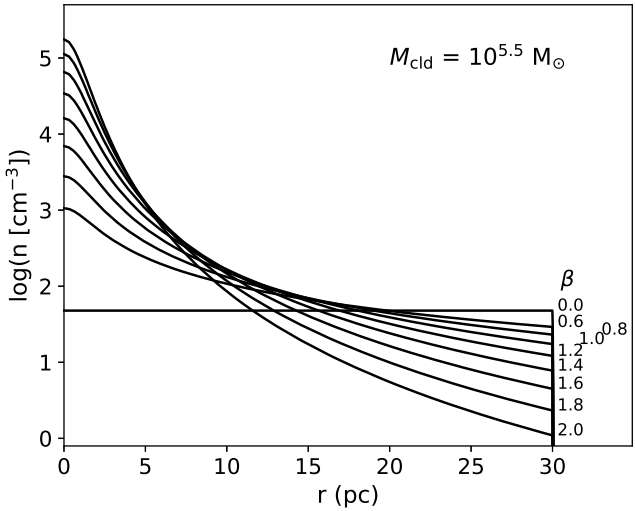}
\caption{\label{beta_prof} Radial profiles of a cloud with $M\ut{cld}=10^{5.5}$ M$_{\odot}$ and $R\ut{cld}=30$ pc, for the various values of $\beta$.}
\end{figure}

We consider a spherically symmetric gaseous cloud with initial density $\rho$ given by a generalized Schuster profile \citep{1998SerAJ.158...15N} 
\begin{equation}
\rho(r) = M\ut{cld} f\ut{Sch}(r, \beta, R\ut{c}, R\ut{cld})\ ,
\end{equation}
\begin{equation}
f_\mathrm{Sch}(r,\beta,R\ut{c},R\ut{cld}) = \frac{C_\mathrm{Sch}(\beta,R\ut{c},R\ut{cld})}{\left[1 + (r/R_c)^2\right]^{\beta}} \quad \mathrm{for} \quad r \le R\ut{cld}
\label{eq:schuster}
\end{equation}
\begin{equation}
C\ut{Sch}(\beta,R\ut{c},R\ut{cld}) = \frac{3}{4\pi} R\ut{cld}^{-3}
\left[{}_{2}F_{1}\left(\frac{3}{2},\beta,\frac{5}{2},-\frac{R\ut{cld}^2}{R\ut{c}^2}\right)\right]^{-1}
\end{equation}
where $r$ is the radial coordinate, $\beta$ is an index indicating the steepness of the profile, $R\ut{c}$ is the core radius, $R_\mathrm{cld}$ is the outer radius of the cloud, and $_{2}F_{1}$ is the hypergeometric function. The normalisation function $C\ut{Sch}(\beta, R\ut{c}, R\ut{cld})$ is obtained by integrating the Schuster profile over the sphere. Fig.~\ref{beta_prof} displays the initial density profile of a cloud with mass $M\ut{cld}=10^{5.5}$\;M$_{\odot}$, radius $R\ut{cld}=30$\;pc, and several values of $\beta$. The initial temperature of the cloud is $T\ut{cld} = 10$\;K, the gas metallicity, $Z$, is either solar or $0.4$ of the solar to match that of the LMC, and the mean molecular weight of the gas particles is $\mu\ut{m} = 2.5$ for $Z = Z_\odot$ or $\mu\ut{m} = 2.48$ for $Z = 0.4 Z_\odot$.

For $r > R\ut{cld}$, the cloud is surrounded by low-density ambient gas with temperature $T\ut{amb}=10^{4}$\;K and such a density that establishes a pressure equilibrium across the cloud surface at $R\ut{cld}$. This ambient medium is considered for purely numerical reasons (to avoid vacuum in hydrodynamic simulations) and it has no impact on results.

\begin{table*}      
\caption{\label{tab:params} Parameters of the performed simulations. The first, second and third sections from the top describe parameters of the cloud, star cluster and the computational domain, respectively.}
\begin{tabular}{c c c} 
\hline
Parameter & Setup A & Setup B \\
\hline
$\log (M\ut{cld}$ [M$_{\odot}$]) & 5.5, 6.0, 6.5 & 4.5, 5.0, 5.5, 6.0, 6.5 \\
$n_0$ [cm$^{-3}$] & 500, 700, 1000, 1600, 2500 & $r-$dependent \\
$\beta$ & 0.0 & 0.6, 0.8, 1.0, 1.2, 1.4, 1.6, 1.8, 2.0\\
$R\ut{cld}$ [pc] & $[3M\ut{cld}/(4\pi n_0 \mu\ut{m} m\ut{H})]^{1/3}$ & 10, 20, 30, 40\\
$R\ut{c}$ [pc] & -- & 1.5\\
$Z$ & $Z_{\odot}$, $0.4 Z_{\odot}$ & $0.4 Z_{\odot}$\\
\hline
SFE$_{0}$ & 0.01, 0.03, 0.05, 0.10, 0.20, 0.30 & 0 \\
$\beta_{*}$ & 1.5 & 1.5 \\
$R_{*}$ [pc] & 1 & 0.15 \\   
 $R_{\text{c},*}$ [pc] & 0.2 & 0.03 \\
\hline
comp. domain radius [pc] & 30 & $R\ut{cld}+10$ \\
number of grid cells & 1024 & 1024 or $2048$ \\

\hline
\end{tabular}
\end{table*}

\subsection{Star cluster}
\label{stars}

The star cluster is always located in the centre of the cloud, i.e. the centre of the coordinate system. Its mass is denoted $M_*$ and it can vary with time. The stellar density is given by the Schuster distribution $f\ut{Sch}(r,\beta_*, R_{\rm c,*}, R_*)$ of the same form but with different parameters than the density of the cloud (see Eq.~\ref{eq:schuster}). For all models we set the slope of the stellar density distribution as $\beta_* = 1.5$, as it approximates well the King density profile of star clusters \citep{1962AJ.....67..471K}. We set the ratio between the star cluster radius and its core radius as $R_* = 5R_{\rm c,*}$, motivated by the observations of the Arches cluster \citep{2009A&A...501..563E}.

The star cluster can be either formed instantaneously in the beginning (we denote this option setup~A), or during the calculation from the gas entering the central region (setup~B). In the former case, the star cluster mass, $M_*\equiv m_{*,0}$, is a free parameter. The main aim of setup~A is to compare the results to the semi-analytic study of a similar model by R18. They parameterize their models by the initial star formation efficiency defined as the ratio between the mass of the star cluster and the cloud, and hence we define the same quantity here as
\begin{equation}
    \mathrm{SFE_0} = m_{*,0} / M\ut{cld} \ .
\end{equation}
In both setup~A and B we use small letters (e.g. $m_{*,0}$) to denote properties of individual stellar populations of the cluster, differing by their age and mass. In setup~B, no stars exists in the beginning of the calculation. Instead, all the gas within a sphere with radius $R_*$ with density higher than $\rho\ut{w,c}$\footnote{Prior to the time when feedback is initiated ($M_*=100$~\MSun{}; see \S\ref{sfb}) this density threshold is set to 10$^{-29}$\,g\,cm$^{-3}$.} is converted into stars at each time step of the hydrodynamic code (see \S\ref{code}). 
Simultaneously, the mass of the formed stars and the time of their formation are recorded. The density threshold, $\rho\ut{w,c}$, is the central (i.e. maximum) density of the star cluster wind resulting from all winds of the stars formed so far (see below for its calculation). This ensures that the star cluster wind is not affected the star formation procedure. Since $\rho\ut{w,c}$ is always much lower that the density of the molecular gas (parts of the cloud) inflowing into the star cluster volume due to the gravity, we can consider that basically all the gas that gets into the central region is converted into stars.

The star cluster consists of a set of stellar populations characterised by their mass, $m_{*,j}$, and formation time, $t_j$, with $j$ being an index of a population. In setup~A, there is only a single population with mass $m_{*,0}$. Alternatively, if the cluster is formed continuously (setup~B), there is one stellar population per each time step of the hydrodynamic code, and the star cluster mass at time $t_j$ is
\begin{equation}
M_{*,j}= \sum_{k=0}^{j} m_{*,k}\\
\end{equation}
where $m_{*,k}$ is the stellar mass formed between times $t_{k}$ and $t_{k+1}$, i.e. during time step $k$.

\begin{figure*}
\centering
\includegraphics[width=\textwidth]{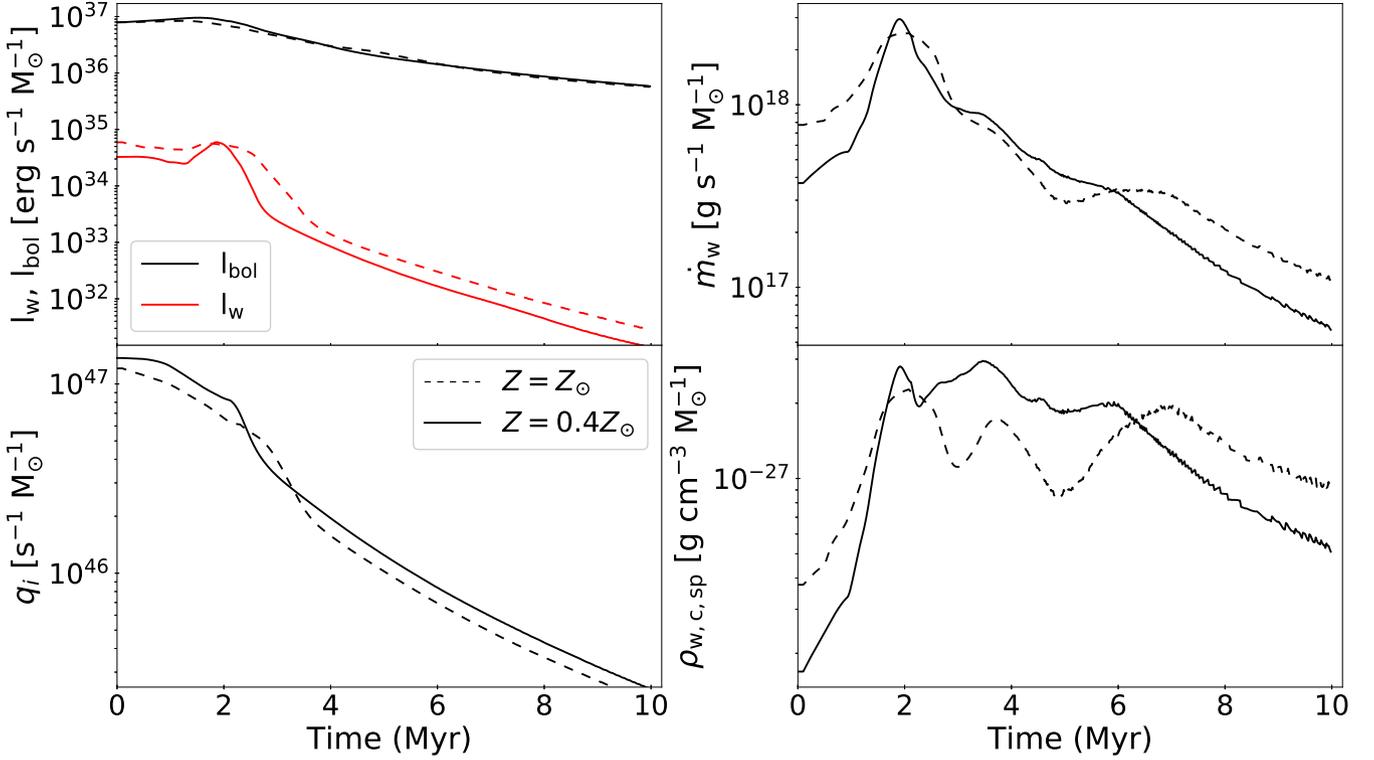}
\caption{Evolutionary properties per unit stellar mass of a single population as function of time. Calculations were made using the stellar models by \citet{Szecsi2022} at solar metallicity (dashed lines) and at $Z=0.4 Z_{\odot}$ (solid lines).}
\label{evol_pop}
\end{figure*}

We assume that each stellar population is formed at a given time with the initial mass function (IMF) given by \cite{2013MNRAS.429.1725M}, which is mathematically convenient approximation of the standard IMF by \cite{2003PASP..115..763C}. Stars are formed with masses in range between $0.08$\;\MSun{} and the minimum between $M_{*}$ and the upper limit of the stellar evolution tracks $500$\;\MSun. Only stars with masses in excess of $9$\;\MSun{} contribute to the feedback. The feedback parameters are calculated by the simple population synthesis code \textsc{synStars} using the Bonn Optimized Stellar Tracks \cite[BoOST;][]{Szecsi2022}. The metallicity of the stars is always the same as the metallicity of the cloud, $Z$, i.e. either $Z_\odot$ or $0.4Z_\odot$. The calculated feedback parameters of a single population are the stellar wind mechanical luminosity, $l_w$, the mass loss rate, $\dot{m}\ut{w}$, the bolometric luminosity, $l\ut{bol}$, and the rate of emitted ionising photons per second, $q\ut{i}$. Fig.~\ref{evol_pop} shows the time evolution of these quantities normalized by the mass of the stellar population. Additionally, we calculate from $l_{w}$ and $\dot{m}_{w}$ the central density of the cluster wind, $\rho\ut{w,c,sp}$, resulting from stars of a single population using the semi-analytic code \textsc{windcalc} \citep{2017ApJ...835...60W} based on the procedure suggested by \citet{2004ApJ...610..226S}.

The collective values of the above quantities for the whole cluster can be obtained by summing up over the populations. If we denote by $x_j$ one of the quantities $l_{w}$, $\dot{m}_{w}$, $l\ut{bol}$, $q\ut{i}$ and $\rho\ut{w,c,sp}$ at time $t_j$, the corresponding quantity $X_j$ (i.e. $L_{w}$, $\dot{M}_{w}$, $L\ut{bol}$, $Q\ut{i}$ or $\rho\ut{w,c}$) for the whole star cluster is
\begin{equation}
X_{j}= \sum_{k=0}^{j} x_{j-k} \times m_{*,k}
\end{equation}
where $x_{j-k}\equiv x(t_{j}-t_{k})$, is the single population quantity with age $t_{j}-t_{k}$, normalized per unit stellar mass.

\subsection{Stellar feedback}
\label{sfb}

Stars of the central cluster provide feedback via their winds, ionising and non-ionising radiation. The stellar winds feedback is mechanical and it is modelled by inserting the gas and the thermal energy into the cluster volume. The ionising radiation ionises and heats the gas within the Str{\"o}mgren radius and both types of radiation generate pressure acting on gas and dust which we assume to be dynamically coupled. All forms of the feedback are activated only if $M_{*} > M\ut{SF} \equiv 100$\;\MSun; the time when it happens is called $t\ut{SF}$. Threshold $M\ut{SF}$ approximately corresponds to the mass of the star cluster containing a single massive star considering the standard IMF.

We follow the model by \citet{1985Natur.317...44C} assuming that individual stellar winds collide and their kinetic energy is thermalized. As a result, the wind material is heated to $10^7-10^8$\;K and the high pressure of this hot gas drives the star cluster wind expanding outwards. It can be simulated \citep[see e.g.][]{2000ApJ...536..896C, 2007ApJ...658.1196T} by inserting the gas and the thermal energy into the star cluster volume, with insertion densities $q\ut{m}$ and $q\ut{e}$, respectively, following the stellar density radial profile,
\begin{gather}
    q\ut{m}(r) = \dot{M}\ut{w}f\ut{Sch}(r, \beta_*, R_{\rm c,*}, R_*) \label{eq:mrate} \\
    q\ut{e}(r) = L\ut{w}f\ut{Sch}(r, \beta_*, R_{\rm c,*}, R_*)  \label{eq:erate}
\end{gather}

Stars of the cluster produce radiation with bolometric luminosity $L\ut{bol}=L\ut{n}+L\ut{i}$, where $L\ut{n}$ and $L\ut{i}$ are luminosities of the non-ionizing ($h\nu < 13.6$\;eV) and ionizing ($h\nu > 13.6$\;eV) radiation, respectively. The mean energy of the ionizing photons is defined as $\langle hv \rangle\ut{i} = L\ut{i}/Q\ut{i}$, where $Q\ut{i}$ is the rate of emitted ionising photons per second. 

Based on the study of \cite{2011ApJ...732..100D} on dusty \ion{H}{II} regions, the radiation pressure force, $dF\ut{rad}$, exerted on an infinitesimally thin spherical shell of radius $r$ with volume $dV=4\pi r^{2}dr$ and number density $n$ is given by
\begin{equation}
    dF\ut{rad} = \frac{L\ut{n}e^{-\tau} + L\ut{i}\phi}{c} n\sigma\ut{d} dr + \alpha_{B}n^{2}\frac{\langle hv \rangle\ut{i}}{c}dV   \label{eq:frad}
\end{equation}
where $\alpha_{B}=2.6\times10^{-13}$ cm$^{3}$ s$^{-1}$ is the recombination coefficient to the excited states and $\sigma\ut{d}$ is the dust cross-section per hydrogen nuclei; it is set to $\sigma\ut{d}=1.5\times10^{-21}$\;cm$^{2}$ for $Z = Z_\odot$ and to $\sigma\ut{d}=6\times10^{-22}$\;cm$^{2}$ for $Z=0.4 Z_\odot$ \citep{2017MNRAS.470.4453R}. We assume that the photons produced by recombinations to the ground state are re-absorbed close to their source \citep[``on-the-spot'' approximation;][]{1974agn..book.....O}. The functions $\tau(r)$ and $\phi(r)$ for the optical depth and the attenuation of the ionizing radiation, respectively, are given by
\begin{gather}
    \frac{d\phi}{dr} = -\frac{1}{Q\ut{i}}\alpha_{B}n^{2}4\pi r^{2} - n\sigma\ut{d}\phi\label{eq:phi}\\
    \frac{d\tau}{dr} = n\sigma\ut{d}\label{eq:tau}
\end{gather}
Equations (\ref{eq:phi}) -- (\ref{eq:tau}) are integrated at each time step of the hydrodynamic code with the boundary conditions $\phi(0)=1$ and $\tau(0)=0$, and functions $\phi(r)$ and $\tau(r)$ are determined. The radius at which $\phi(r)$ drops to zero is identified, denoting the radius of the ionization front, $R\ut{IF}$. 

Equation~\eqref{eq:frad} is then used to calculate the amount of momentum deposited into each grid cell at each time step. Its first right-hand-side (rhs) term represents the radiation pressure of both ionising and non-ionising radiation acting on the dust. The second rhs term describes the momentum deposited by the ionising radiation to the gas, and it is applied only in cells where $\phi(r) > 0$, i.e. for $r < R\ut{IF}$. Additionally, the gas at $r < R\ut{IF}$ is ionised, i.e. its mean molecular weight is set to $\mu\ut{i}$ corresponding to the ionized gas, and heated, i.e. its temperature is set to $10^4$\;K for gas that is cooler than that. Heating by the non-ionising radiation is neglected. Similarly, the reprocessed infrared radiation from dust grains is also ignored, as it is significant only in dust-enriched environments \citep{2015ApJ...809..187S}, and in general, is shown to be unable to regulate the star formation processes \citep{2022MNRAS.517.1313M}.

\subsection{Simulation setups}

In the first set of simulations (setup~A), the cloud is uniform ($\beta = 0$) and the star cluster exists from the beginning. We perform 180 simulations varying the four following parameters (see Table~\ref{tab:params}): the cloud mass, $M\ut{cld} = 10^{5.5}$, $10^6$, and $10^{6.5}$\;\MSun, the initial cloud particle density, $n_0 = \rho_0 / \mu\ut{m}m\ut{H} = 500, 700, 1000, 1600,$ and $2500$\;cm$^{-3}$ where $m\ut{H}$ is the proton mass, the gas and stars metallicity $Z = 0.4 Z_{\odot}$ and $Z_{\odot}$, and the initial star formation efficiency, $\mathrm{SFE}_{0} = 0.01, 0.03, 0.05, 0.1, 0.2,$ and $0.3$ that defines the mass of the star cluster, $M_*$. The star cluster radius is always $R_* = 1$\;pc. The selected parameters follow closely those used in R18 to make the comparison straightforward. The computational domain has radius $R\ut{cd} = 30$\;pc, and consists of $1024$ grid cells.

In setup~B, the star cluster is formed during the simulation. Uniform clouds (and clouds with small slopes $\beta$) have relatively low gas density in their centres leading to the slow inflow of the gas there. As a result, the star cluster mass grows slowly, and the feedback is unable to stop the collapse until the vast majority of the cloud enters the star cluster volume where it is converted to stars. This is clearly a result of the artificial symmetry of the model and as such it is not astrophysically interesting. Therefore, we focus on density profiles with $\beta=0.6-2.0$, at a step of $0.2$\;dex. Additionally, we vary the cloud mass, $M\ut{cld} = 10^{4.5}$, $10^{5}$, $10^{5.5}$, $10^{6}$ and $10^{6.5}$\;\MSun, and the cloud radius, $R\ut{cld} = 10$, $20$, $30$ and $40$\;pc. The cloud core radius is always $R\ut{c} = 1.5$\;pc. Altogether, we perform $160$ simulations under setup~B (see Table~\ref{tab:params}). The star cluster radius is $R_* = 0.15$\;pc, chosen to be sufficiently small in order to minimize any contribution to the star formation process by the enclosed gas at $t=0$. The computational domain has radius $R\ut{cd} = R\ut{cld} + 10$\;pc and it consists of $1024$ grid cells for simulations with $R\ut{cld} < 30$\;pc or of $2048$ grid cells, otherwise.

Simulations terminate either when the entire cloud has collapsed into the center (SFE = 1), or when, fraction of the gas has been converted into stars and the size of the \ion{H}{II} region has reached that of the computational volume. The latter condition indicates that stellar feedback has driven the shell of the residual swept-up cloud out of the explored domain.


\section{Numerical code}
\label{code}

To calculate the model evolution, we use a numerical code based on the high-performance parallelized hydrodynamic framework \textsc{flash} \citep{2000ApJS..131..273F}. We configure \textsc{Flash} to solve the one-dimensional Euler equations for compressible flow in spherical coordinates in the following form 
\begin{gather}
    \frac{\partial \rho}{\partial t}+\frac{1}{r^2}\frac{\partial}{\partial r}(r^2 \rho u)=q_{m} \label{eq:hydro1}\\
    \frac{\partial \rho u}{\partial t}+\frac{1}{r^2}\frac{\partial}{\partial r}(r^2 \rho u^2)+\frac{\partial P}{\partial r}=\rho g-q_{m}u-f\ut{rad} \label{eq:hydro2}\\
    \frac{\partial \rho E}{\partial t}+\frac{1}{r^2}\frac{\partial}{\partial r}(r^2 (\rho E + P) u)]=\rho ug + q_{e} + q\ut{rad} - q_{c} \label{eq:hydro3}
\end{gather}

using the Piecewise Parabolic Method \citep[PPM;][]{1984JCoPh..54..174C} 
with a time step $\Delta t$ controlled by the Courant$-$Friedrichs$-$Levy criterion. In the above equations, $t$ is the time, and $\rho$, $u$, and $E$ are the gas density, velocity and the total energy per unit mass, respectively. Quantity $E$ is the sum of the internal energy $\varepsilon$ and the kinetic energy per unit mass, i.e.
\begin{equation}
      E = \varepsilon + \frac{u^2}{2}
\end{equation}
where $\varepsilon = P/[\rho(\gamma - 1)]$ with $\gamma$ being the ratio of specific heats and $P$ being the gas pressure given by the ideal gas equation of state
\begin{equation}
    P=\frac{\rho \kappa\ut{B} T}{\mu m\ut{H}}
\end{equation}
where $\kappa\ut{B}$ is the Boltzmann constant. The mean molecular weight $\mu$ depends on the gas metallicity, $Z$ and temperature, $T$. For simplicity, we define three values: $\mu_i$ for the ionised gas with $T > 9700$\;K, $\mu_a$ for the atomic gas with $9300\;\mathrm{K}>T>700\;\mathrm{K}$, and $\mu_m$ for the molecular gas with $T < 300$\;K. Values of $\mu$ in transition regions $300 - 700$\;K and $9300-9700$\;K are interpolated linearly. The actual values of $\mu_i$, $\mu_a$ and $\mu_m$ are calculated from the abundances of species given by $Z$ and the temperature dependent number of free electrons given by \citet[][see their Tab.~2]{2009A&A...508..751S}. The species abundances are taken from \citet{2009ARA&A..47..481A} for $Z=Z_{\odot}$ and scaled appropriately for other $Z$ values. In this work, we use either the metallicity of the LMC ($Z = 0.4 Z_\odot$) or the solar metallicity ($Z = Z_{\odot}$). The corresponding mean molecular weights are $\mu_m = 2.48$, $\mu_a = 1.24$, $\mu_i =  0.597$ for $Z = 0.4 Z_\odot$; and $\mu_m = 2.5$, $\mu_a = 1.25$, $\mu_i = 0.601$ for  $Z = Z_{\odot}$.

The star cluster wind is modelled by inserting the mass and energy into the star cluster volume ($r < R_*$) at rate densities $q_{m}$ and $q_{e}$ given by Equations \ref{eq:mrate} and \ref{eq:erate}, respectively. The momentum in the affected cells is conserved during the insertion procedure \citep[see][]{2008ApJ...683..683W}.

The gravitational acceleration, $g$, consists of two components; one for the self-gravity of the gas, $g\ut{gas}$, and one for describing the attractive force between the star cluster, $g_*$. Both accelerations, at a given coordinate $r$, are obtained simply from the mass, $M(r),$ enclosed within the sphere of radius $r$
\begin{equation}
  g\ut{gas} = -GM\ut{gas}(r)r^{-2}, \qquad
  g_* = -GM_*(r)r^{-2}
\end{equation}
where $G$ is the constant of gravity. For the star cluster, the enclosed mass is 
\begin{equation}
  M_*(r) = \left\{
  \begin{array}{lll}
  M_*C\ut{Sch}(\beta_*,R_{\rm c,*},r) & \mathrm{for} & r \le R_* \\
  M_* & \mathrm{for} & r > R_*
  \end{array}
  \right.
\end{equation}
The enclosed mass of the gas, $M\ut{gas}(r)$, is calculated at each time step by a fast integrator described below.

Determining the gravitational acceleration due to gas self-gravity and solving radiative transport equations (\ref{eq:phi}) and (\ref{eq:tau}) leads to the same mathematical problem -- calculating an integral of a certain power of the particle density, $n$, from the centre to each grid cell at each time step. For that, we develop a fast parallel integrator calculating quantity $y_j$ for each grid cell $j$
\begin{equation}
    y_{j} = \sum_{i=0}^{j} n_{i}^{p} \Delta x_{i} 
\end{equation}
where $n_i$ is the particle density in grid cell $i$, $p$ is an integer number, and $\Delta x_i$ is either the radial extent of the $i$-th grid cell $\Delta r_i$ or its volume $\Delta V_i$. The integrator utilizes a property of the {\sc Flash} domain decomposition algorithm guaranteeing that sub-domains assigned to individual processors are contiguous. This significantly simplifies the communication, as only the sums over the whole sub-domain have to be sent to other processors. The enclosed mass, $M\ut{gas}(r)$, is then calculated by setting $p=1$ and $\Delta x_i = \Delta V_i$; the first rhs term of Eq.~\ref{eq:phi} by setting $p=2$ and $\Delta x_i = \Delta V_i$; and the first rhs term of Eq.~\ref{eq:tau} by setting $p=1$ and $\Delta x_i = \Delta r_i$. The calculation of the second rhs term of Eq.~\ref{eq:phi} uses $\Delta x_i = \Delta r_i$, and additionally adopts $\phi_i$ from the previous hydrodynamic time step.

The gas looses energy by emitting the radiation at rate
\begin{equation}
    q_{c}=- n\ut{H}^2 \Lambda(T, Z)    
\end{equation}
where $n\ut{H}$ is the mean mass per the Hydrogen nuclei and $\Lambda(T, Z)$ is the cooling function by \cite{2009A&A...508..751S}. They tabulate it for each specie individually making it convenient to scale it for any metallicity. The energy losses are assumed to be optically thin, i.e. the energy is subtracted from each grid cell and it is not followed further. To account to the ionising radiation heating, we use a simple approximation of the constant \ion{H}{II} region temperature: the temperature in grid cells where $\phi > 0$ (intensity of the ionising radiation is non-zero) is not allowed to drop below $10^4$\;K. In grid cells where $\phi = 0$, a constant heating is used with rate $q\ut{rad} = n\ut{H}\Gamma$ with $\Gamma = 2\times 10^{-26}$\;erg\,s$^{-1}$ taking into account several common heating processes in the interstellar medium \citep{2002ApJ...564L..97K}.

The acceleration due to the radiation pressure in Eq.~\ref{eq:hydro2}, $f\ut{rad}=dF\ut{rad}/dV$, is given by Eq.~\ref{eq:frad}. It can be easily determined for each grid cell knowing functions $\phi$ and $\tau$.

We investigate the sensitivity of the model evolution on the grid resolution. For our test purposes, we run setup~B simulations of five different cloud models using three resolutions (grid cell numbers). The results are provided in the appendix (Fig. \ref{resol_app}) and they show that the differences among the calculations at different resolution are smaller than $\sim0.1$ dex for all five models.

\section{Results}
\label{results}

The results of the two simulation sets are reported below. Common features of the model evolution are described in \S\ref{shell}, followed by \S\ref{sec:setupA} and \S\ref{sec:setupB} discussing results of the setup~A and setup~B runs, respectively.

\subsection{General model evolution}
\label{shell} 

\begin{figure*}
\centering
\includegraphics[width=\textwidth]{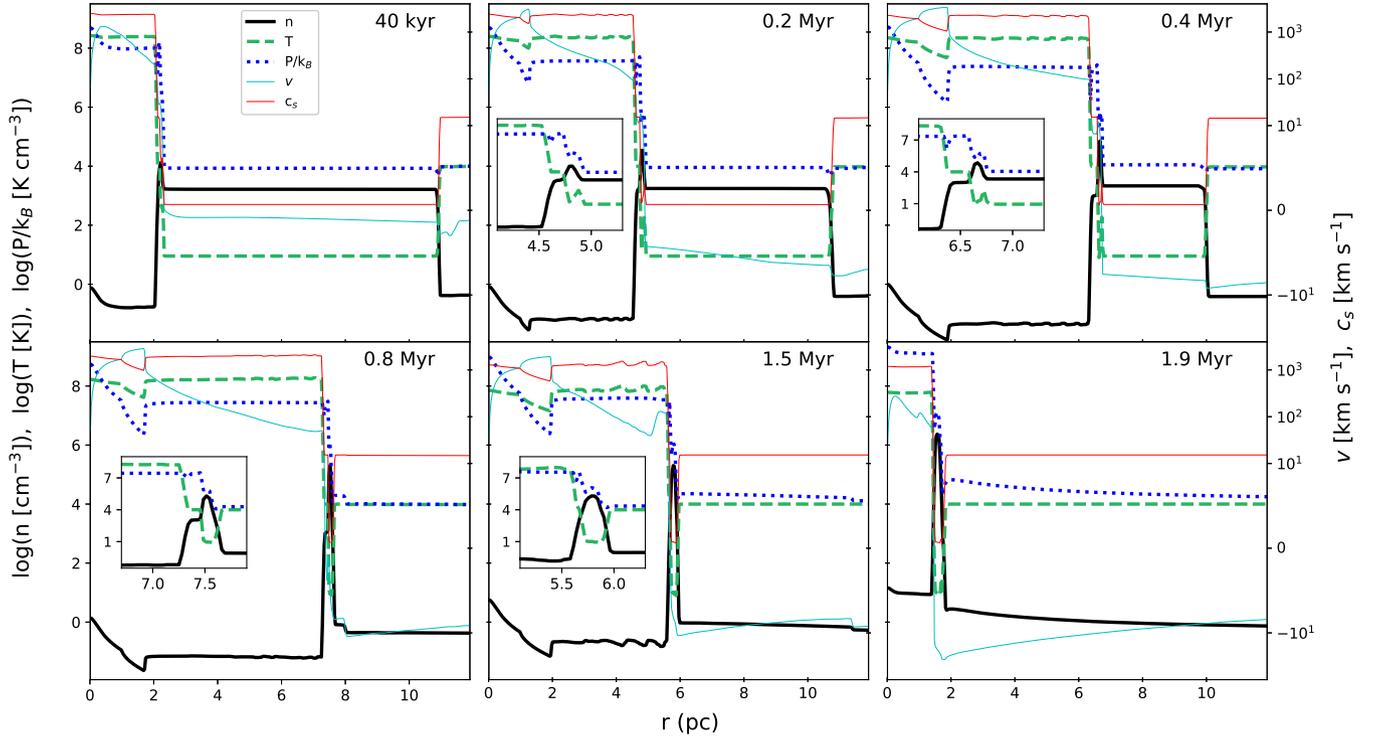}
\caption{\label{shell_evol} Radial profile at different times of the various properties of the gas; particle density (black thick line), temperature (green dashed line), and pressure (blue dotted line) in log scale (left axis), as well as velocity (cyan thin line) and sound speed (red thin line) in km~s$^{-1}$ (right axis). The simulations follow setup~A assuming a cloud with $M\ut{cld}=10^{5.5}$ M$_{\odot}$ and $n=1000$ cm$^{-3}$, which collapses against the feedback from a central cluster that is formed with SFE$_{0}=0.03$. A zoomed view of the shell is displayed by insets within the panels for $t=0.2-1.5$ Myr. For simplicity, only the particle density, temperature and pressure are shown in the insets.}
\end{figure*}

The interaction of the star cluster wind and radiation with the cloud leads to the establishment of a bubble described by \citet{1975ApJ...200L.107C, 1977ApJ...218..377W}. Fig.~\ref{shell_evol} shows its structure and evolution for a setup~A model with parameters $M\ut{cld} = 10^{5.5}$\;\MSun, $n_0 = 1000$\;cm$^{-3}$ and SFE$_{0}=0.03$ (i.e. $M\ut{*} \simeq 10^4$\;\MSun). The components of the bubble are well visible e.g. at $0.4$\;Myr (see the top right panel of Fig.~\ref{shell_evol}, and the inset zooming in on the \ion{H}{II} region and the cold shell). The wind is formed within the star cluster, and at its edge the wind velocity crosses the sound speed (see cyan and red thin lines, cnf. \citealt{1985Natur.317...44C}). The wind expands freely, up to the reverse shock, $R\ut{RS} \simeq 2$\;pc where its kinetic energy is thermalized and the gas is heated to $\sim 10^8$\;K. The shocked wind region extends up to the contact discontinuity at $R\ut{CD} \simeq 6.4$\;pc, where it adjoins to the cloud gas ionised by the stellar radiation (\ion{H}{II} region; see the drop of the temperature, green dash-dotted line, from $10^8$ to $10^4$\;K). The \ion{H}{II} region is relatively thin as it is compressed by the high pressure of the shocked wind. Its outer edge is at the ionisation front $R\ut{IF} \simeq 6.6$\;pc, where the high energy photons ($hv>13.6$) are depleted. The \ion{H}{II} region is surrounded by the thin shell of cold gas composed of the accreted cloud material. The temperature of the shell is close to $10$\;K, however, its outer edge at the shock front, $R\ut{SF} \simeq 6.7$\;pc, is slightly heated by the compression there. The original cloud is located at even higher radii, falling inwards onto the shell due to the gravity (see the cyan line of the gas velocity in Fig.~\ref{shell_evol}). The outermost part of the computational domain is filled by the rarefied warm ambient medium.

In setup~A runs, the cluster is inserted into the simulation in the beginning, and it is always massive enough to form the bubble almost immediately. In setup~B runs, it may take a certain time until the cluster mass reaches the value when the feedback is powerful enough to reverse the cloud collapse. We define time $t\ut{e}$ when the thermal pressure in the central region exceeds the ram pressure of the infalling cloud and the bubble starts to expand. Furthermore, the inflow of gas into the star cluster volume and the growth of $M_{*}$ stop at $t\ut{e}$.


Once the bubble is formed, it expands and the dense cold shell on the bubble surface accretes the cloud mass. Simultaneously, the cloud collapses adding more mass to the shell. This process can lead to two outcomes: either the cloud becomes unbound and the expansion continues indefinitely, or the expansion stops at some point and the shell re-collapses into the centre where it is immediately converted into stars. Fig.~\ref{shell_evol} shows the latter case, with the shell starting to collapse at $\sim 0.7$\;Myr and reaching the centre shortly before $2$\;Myr. The time evolution of the ionisation front, $R\ut{IF}$, can be seen also in Fig.~\ref{evol_stroe} (middle row, right column, yellow solid line).

\subsection{Models with pre-existing star cluster (setup~A)}
\label{sec:setupA}

\begin{figure*}
\centering
\includegraphics[width=\textwidth]{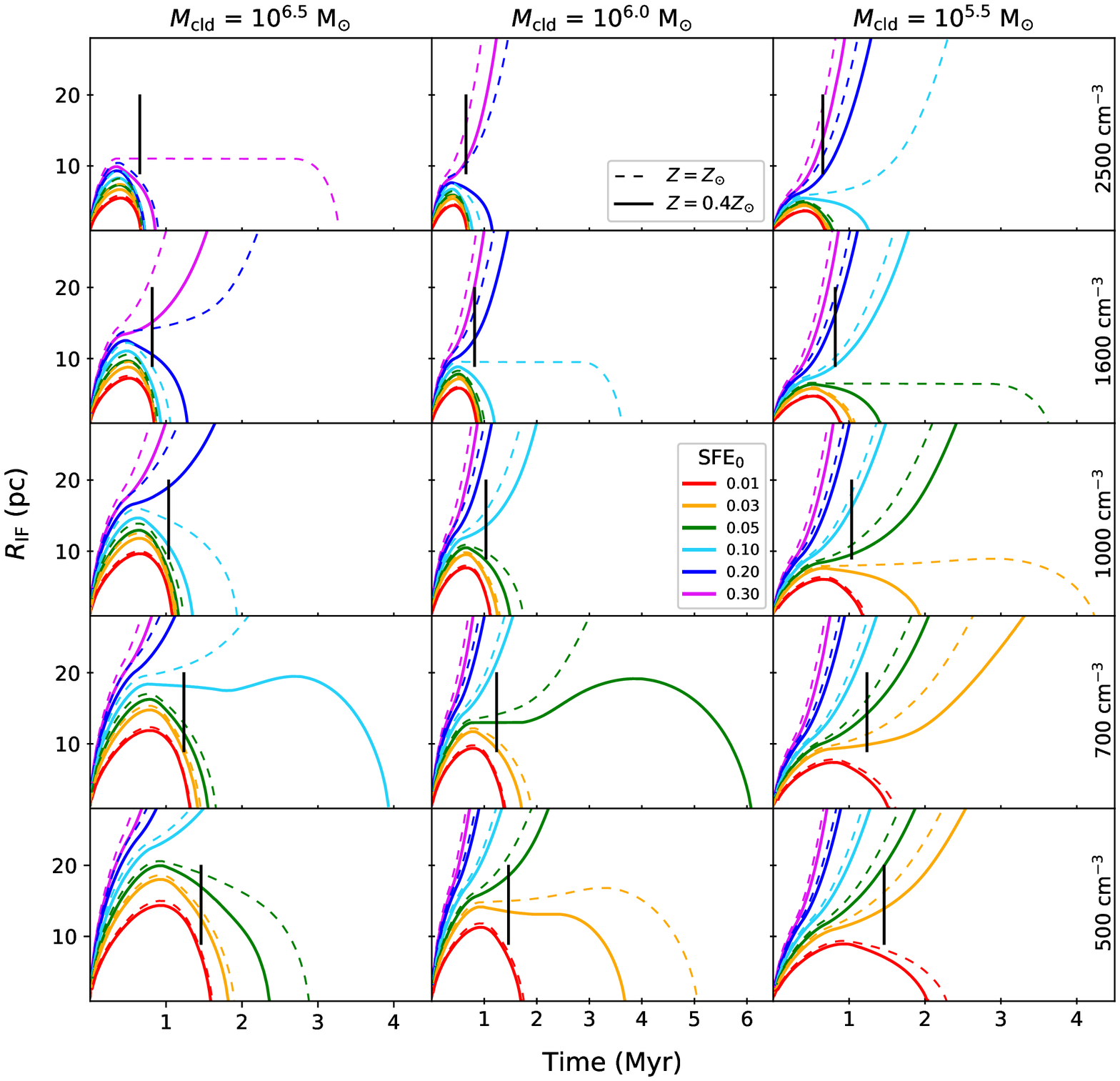}
\caption{\label{evol_stroe} Evolution of the outer radius of the \ion{H}{II} region as function of the mass and density of the cloud. The plotted lines are color coded with respect to SFE$_{0}$. Dashed and solid lines show models $Z = Z_{\odot}$ and $Z = 0.4 Z_{\odot}$, respectively. In each panel, $\tau\ut{ff}$ is indicated by a vertical line.}
\end{figure*}

In Fig. \ref{evol_stroe}, we display the evolution of the outer radius of the \ion{H}{II} region, $R\ut{IF}$, for all setup~A models. Depending on the initial parameters $M\ut{cld}$, $n_0$ and $\mathrm{SFE_0}$, there are two qualitatively different types of behaviour. With high $M\ut{cld}$, high $n_0$ and low $\mathrm{SFE_0}$, the bubble expands to a certain radius and then re-collapses. In the opposite case, i.e. low $M\ut{cld}$, low $n_0$ and high $\mathrm{SFE_0}$, the bubble expansion continues indefinitely and the cloud is unbound. Considering the extreme cases, in the most massive and compact cloud with $M\ut{cld} = 10^{6.5}$\;\MSun{} and $n_0 = 2500$\;cm$^{-3}$, the bubble re-collapses even with the highest $\mathrm{SFE_0} = 0.3$ corresponding to the star cluster with mass $M_{*,0} \simeq 10^6$\;\MSun{} (see the top left corner of Fig.~\ref{evol_stroe}). On the other hand, the cloud with $M\ut{cld} = 10^{5.5}$\;\MSun{} and $n_0 = 500$\;cm$^{-3}$ is unbound with all models with $\mathrm{SFE_0}\geq0.03$ ($M_{*,0} \simeq 10^4$\;\MSun; bottom right corner of Fig.~\ref{evol_stroe}).


Time at which the bubble in bound models re-collapses is given by the cloud free-fall time
\begin{equation}
    \tau\ut{ff} = \left(\frac{3\pi}{32 G\rho_0}\right)^{1/2} \simeq 1.03 \left( \frac{n_0}{10^3~\mathrm{cm}^{-3}} \right)^{-1/2} \mathrm{Myr} \ .   
\end{equation}
In models with low $\mathrm{SFE_0}$, where the impact of the feedback from the star cluster is not significantly affecting the overall cloud collapse, the bubble collapses at almost exactly $\tau\ut{ff}$. The bubble re-collapse takes longer for models with higher $\mathrm{SFE_0}$. In several cases with $\mathrm{SFE_0}$ close to the value needed to unbound the cloud we observe periods when $R\ut{IF}$ stays almost constant for several Myr due to the equilibrium between the thermal pressure of the bubble interior and the ram pressure of the infalling cloud. This resembles behaviour described in \citet{2003A&A...411..397T}. We stress, however, that the exclusion of supernova feedback from our model renders the outcome beyond $t \sim 3.5$\;Myr uncertain.


To explore the impact of metallicity on the bubble evolution, we calculate all setup~A models with two values $Z = Z_{*} = Z_{\odot}$ and $Z = Z_{*} = 0.4 Z_{\odot}$, shown in Fig.~\ref{evol_stroe} by dashed and solid lines, respectively. On one hand, the mechanical luminosity of stellar winds, $L_w$, is higher with higher $Z_{*}$ due to the wind line-driven nature, making the feedback more efficient. On the other hand, the bolometric luminosity of the star cluster is slightly lower with higher $Z_{*}$ making the radiation pressure feedback less efficient. Additionally, higher gas metallicity, $Z$, results in higher cooling of the hot gas reducing its temperature and hence the thermal pressure. Fig.~\ref{evol_stroe} shows that the effect of higher $L_w$ dominates in all cases and the bubble expands faster with higher $Z$ and $Z_{*}$. Essentially thus, our findings contradict those by studies that suggest low metallicity to be supportive for the growth of \ion{H}{II} regions and the reduction of SFE, however ignoring the impact of stellar winds \cite[e.g.][]{2020MNRAS.497.3830F,2021MNRAS.501.4136A}.


\begin{figure*}
\centering
\includegraphics[width=\textwidth]{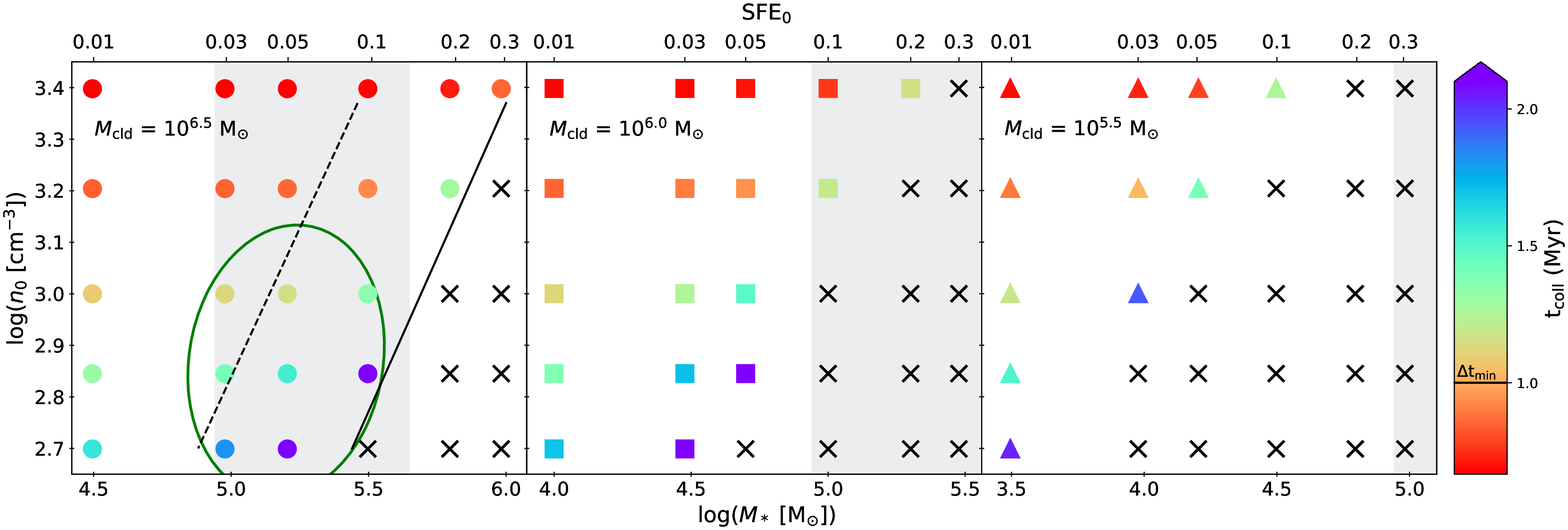}
\caption{Collapsing times of the shell as function of cloud density and SFE$_{0}$. The ``X'' markers stand for simulations that predict a shell expansion. For a cloud with $M\ut{cld}=10^{6.5}$ M$_{\odot}$, the dashed line indicates the threshold SFE$_{0}$ from \protect\cite{2018MNRAS.473L..11R} that differentiates collapsing models from those with a disruptive feedback. Our respective threshold (solid line) is shown to be $3-4$ times higher. The mass of NGC 2070 as source of feedback at $t=0$ is displayed with a grey shaded area. An horizontal line in the colorbar indicates the minimum age separation between NGC 2070 and R136, the latter being suggested to have formed upon re-collapse of the gas. For $M\ut{cld}=10^{6.5}$ M$_{\odot}$, the models consistent with the re-collapse scenario are enclosed by a green ellipse.}
\label{tcoll}
\end{figure*}

In Fig. \ref{tcoll}, we display time $t\ut{coll}$ during which the bubble re-collapses as a function of $\mathrm{SFE_{0}}$ and $n_0$, for runs with $Z=0.4 Z_{\odot}$. In the left panel showing the cloud with $M\ut{cld}=10^{6.5}$ M$_{\odot}$, a dashed line indicates the threshold found by R18 to separate models where the cloud is bound and unbound. Our results are qualitatively similar, however, the threshold (shown by solid line in Fig.~\ref{tcoll}) is shifted to higher $\mathrm{SFE_0}$ or lower $n_0$. To unbound the cloud, either $\mathrm{SFE_0}$ higher by factor of $3-4$, or $n_0$ lower by factor $\sim 5$ is needed in comparison to R18. The main reason for this difference is that the self-gravity of the cloud is included in our model, however, it is not in the R18 model (we remind that R18 include the self-gravity of the shell, however, the cloud is assumed to be in virial equilibrium, i.e. stabilised against the gravitational force by the turbulent and/or magnetic pressure). 

Following R18, we apply the model of the shell re-collapse to explain the origin of NGC~2070 and its younger central star cluster R136. NGC~2070 corresponds to the initial star cluster, and for its mass we adopt range $8.7 \times 10^{4}-4.5 \times 10^{5}$\;\MSun \citep{2009AJ....137.3437B, 2015ApJ...811...76C}, shown as gray-shaded area in Fig.~\ref{tcoll}. Another constrain is given by the age difference between NGC~2070 and R136, which is at least $1$\;Myr. It corresponds to time $t\ut{coll}$ of the shell re-collapse and it is shown by color in Fig.~\ref{tcoll}. With cloud mass $10^{6.5}$\;\MSun, several runs with $\mathrm{SFE_0} = 0.03 - 0.1$ and $n_0 \lesssim 1000$\;cm$^{-3}$ fulfill both constrains and are thus consistent with the observations. Clouds with lower masses need higher $\mathrm{SFE_0}$ and higher $n_0$ to justify the above discussed scenario. 



\begin{figure}
\includegraphics[width=8.58cm]{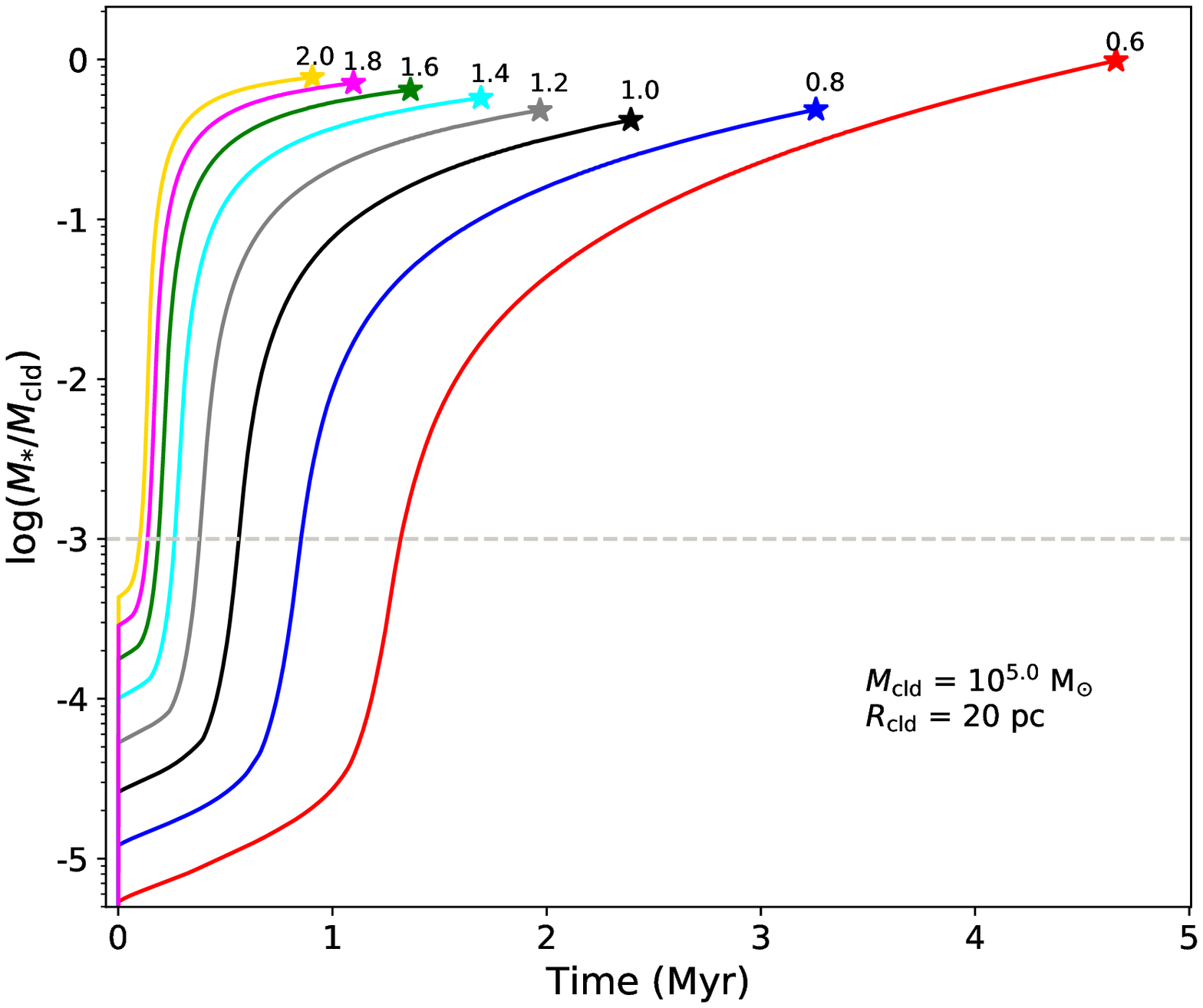}
\caption{\label{sf_rate} Evolution of the stellar mass relative to the mass of the cloud, for the different profiles of gas density. The mass and radius of the cloud at $t=0$ are here set as $M\ut{cld}=10^{5}$ M$_{\odot}$ and $R\ut{cld}=20$ pc. For each of the models, we indicate the $\beta$ value of the cloud steepness, and mark the point (asterisk) when the stellar feedback begins to disperse the residual gas. At the latter point, the mass of the cluster  defines the SFE of the model. A dashed horizontal line indicates the 100 M$_{\odot}$ threshold in the stellar mass for initiating feedback in our simulations. Calculations over the entire set of cloud parameters are shown in the appendix (Fig. \ref{sfr_app}).}
\end{figure}

\subsection{Models with continuous star formation (setup~B)}
\label{sec:setupB}

In setup~B runs, the star cluster is formed during the simulation from the gas that enters the central sphere with radius $R_{*} = 0.15$\;pc. Fig.~\ref{sf_rate} shows time evolution of the star cluster mass, $M_{*}$, for a cloud with initial mass and radius $M\ut{cld} = 10^{5}$\;\MSun and $R\ut{cld} = 20$\;pc, respectively, and for different density slopes $\beta$. All the runs show the qualitatively same behaviour. Initially, $M_{*}$ grows relatively slowly as the cloud starts to contract due to the self-gravity. At a certain time, the mass of the cluster starts to contribute significantly to the gravitational acceleration and the $M_{*}$ growth rate suddenly increases. Shortly after that, $M_{*}$ exceeds threshold $100$\,\MSun and the feedback is switched on. Then, $M_{*}$ continues to grow with decreasing rate until time $t\ut{e}$ (denoted by asterisk in Fig.~\ref{sf_rate}) when the feedback is strong enough to create an expanding bubble and to stop the inflow of the gas into the star cluster volume. Finally, the bubble expands until it reaches the computational domain boundary ending the simulation. All setup~B models exhibit only a single period of star formation; once the star cluster mass was stopped at $t\ut{e}$, the feedback was always strong enough to unbound the cloud. That allows us to define the star formation efficiency, SFE\footnote{Not to be confused with $\mathrm{SFE}_0$, which is a free parameter setting the mass of the pre-existing star cluster in setup~A models.},  as the ratio of the star cluster mass (at $t\ut{e}$ when it is definitely formed) and the initial cloud mass $M\ut{cld}$
\begin{equation}
    \text{SFE} = M_{*}(t_{e})/ M\ut{cld} \ .
\end{equation}


With increasing the steepness of the cloud density profile, $\beta$, the $M_{*}$ grows faster due to the higher rate at which gas flows into the star cluster volume. Additionally, higher $\beta$ results in the higher density of the gas surrounding the star cluster leading to the suppression of the feedback from more massive star clusters. As a consequence, $\mathrm{SFE}$ rises with increasing $\beta$ in the range $\beta = 0.8 - 2.0$. On the other hand, $M_{*}$ grows slowly in clouds with $\beta < 0.8$, and lower $\beta$ results in longer time before the feedback is able to create the bubble and in higher fraction of the cloud being converted into stars. In other words, high SFE is reached either with high $\beta$ due to the fast growth of $M_{*}$ in the beginning and the high central pressure, or with low $\beta$ due to the slow $M_{*}$ growth allowing the cloud collapse to proceed to the advanced stage. Intermediate values of $\beta$ ($\sim 0.8 - 1.0$) lead to the lowest SFE. We denote $\beta\ut{min}$ the slope for which SFE reaches minimum value for given $M\ut{cld}$ and $R\ut{cld}$; the corresponding star formation efficiency is $\mathrm{SFE}(\beta\ut{min})$. The time evolutions of $M_{*}$ for the entire set of setup~B runs are provided in the appendix (Fig. \ref{sfr_app}).

\begin{figure}
\includegraphics[width=\columnwidth]{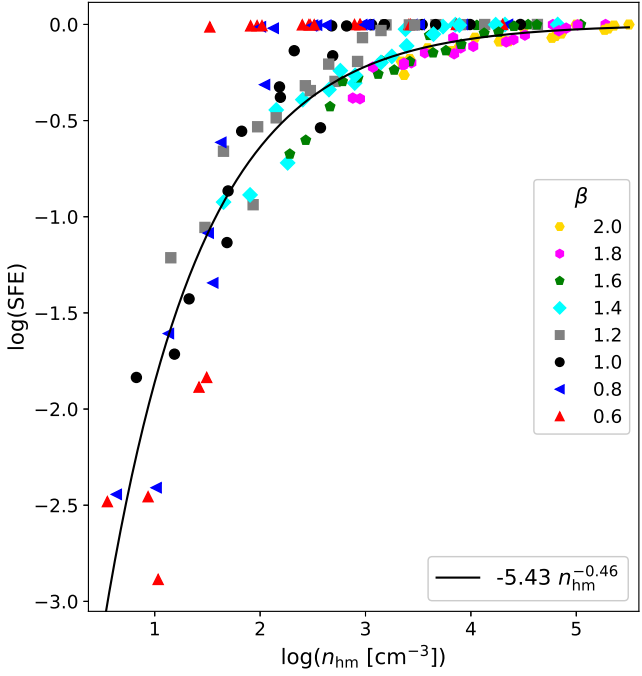}
\caption{\label{sfe_rho} Logarithm of the star formation efficiency (SFE) as a function of the mean particle density within the half-mass radius of the cloud, for the all setup~B models. The color of symbols represents $\beta$. The best fit to the data (disregarding points with $\beta<1$ that show $\log(\mathrm{SFE}) \simeq 0$) is displayed with a solid black line (Eq. \ref{eq:sfe_rho}).}
\end{figure}

We search for a single parameter combining $M\ut{cld}$, $R\ut{cld}$ and $\beta$ giving the tightest relation with SFE, and found that the mean particle density within the cloud half-mass  radius works well in this respect. It is defined as
\begin{equation}
    n\ut{hm} = \frac{3M\ut{cld}}{8 \pi \mu_{c} m\ut{H} R\ut{hm}^3} \ , 
\end{equation}
where $R\ut{hm}$ is the numerically determined half-mass radius of the cloud at time $t\ut{SF}$ when the stellar feedback is activated, i.e. $M_*$ reaches $M\ut{SF}$. Fig.~\ref{sfe_rho} shows SFE as a function of $n\ut{hm}$ for all setup~B models. The color and shape of symbols represent the initial cloud density slope $\beta$. It can be seen that all the models can be divided into two groups. First, there are models with SFE nearly $100$\,\% (i.e. $\log(\mathrm{SFE}) \simeq 0$) with typically $\beta<1$. They are models where it takes a long time until $M_*$ reaches $M\ut{SF}$, and when it finally happens, the cloud collapse proceeded so much that the feedback bubble cannot be formed unless the vast majority of the cloud is converted into stars. Second, the majority of models lie along the line given by the relation
\begin{equation}
    \log(\text{SFE}) = -5.43 \times \left(\frac{n\ut{hm}}{\mathrm{cm}^{-3}}\right)^{-0.46} \ .
    \label{eq:sfe_rho}
\end{equation}
We found the second group astrophysically more interesting, because the volume density in real molecular clouds varies significantly more than in our model, and the situation that the feedback bubble cannot be created before the whole cloud collapses (as in the first group) is highly improbable. 



\begin{figure}
\includegraphics[width=\columnwidth]{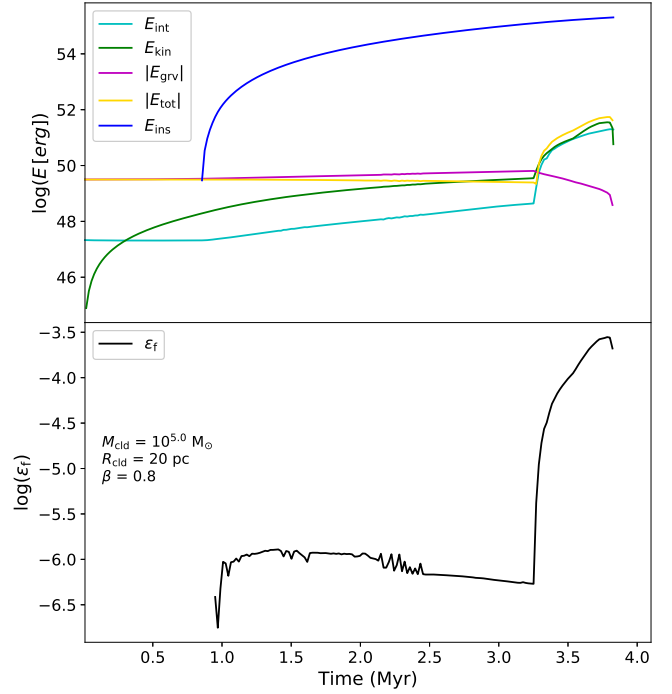}
\caption{\label{fig:energies} \textbf{Top:} Time evolution of the logarithm of the cumulative energy inserted by the all different forms of feedback, $E\ut{ins}(t)$ (Eq.~\ref{eq:Eins}, blue line), and logarithms of components of the gas energy: the internal energy $E\ut{int}(t)$ (cyan), the kinetic energy $E\ut{kin}(t)$ (green), the gravitational energy $E\ut{grv}(t)$ (magenta, Eq.~\ref{eq:Egrv}), and the total energy $E\ut{tot}(t)$ (yellow). \textbf{Bottom:} Time evolution of the logarithm of the instantaneous feedback efficiency $\epsilon\ut{f}(t)$ (see Eq.~\ref{eq:epsf}). All lines in both panels represent a model with $M\ut{cld} = 10^{5}$\,\MSun, $R\ut{cld} = 20$\,pc and $\beta = 0.8$.}
\end{figure}

Fig.~\ref{fig:energies} compares the energy of the feedback to the energy retained by the gas, for a model with $M\ut{cld} = 10^5$\,\MSun, $R\ut{cld} = 20$\,pc and $\beta = 0.8$. The blue line in the top panel represents the cumulative energy inserted by the all forms of the feedback:
\begin{equation}
    E\ut{ins}(t) = \int_{t\ut{SF}}^t [L\ut{bol}(t') + L\ut{w}(t')]dt' \ .
    \label{eq:Eins}
\end{equation}
The other lines are components of the energy associated with gas at given time, specifically, the internal energy $E\ut{int}$ (cyan), the kinetic energy $E\ut{kin}$ (green), and the gravitational binding energy (magenta)
\begin{equation}
    E\ut{grv}(t) = -G \int_0^{R\ut{cd}} \frac{[M\ut{cld}(r,t) + M_*(r,t)]}{r} \rho(r,t) 4\pi r^2 dr
    \label{eq:Egrv}
\end{equation}
where $M\ut{cld}(r,t)$ and $M_*(r,t)$ are fractions of the cloud mass and the cluster mass below radius $r$ at time $t$, respectively, and $\rho(r,t)$ is the gas density. The total energy of the gas is $E\ut{tot}(t) = E\ut{int}(t) + E\ut{kin}(t) + E\ut{grv}(t)$. 

The energy retained by the gas can be estimated by subtracting the initial total gas energy $E\ut{tot}(0)$ from $E\ut{tot}(t)$. This allows us to define the instantaneous feedback efficiency
\begin{equation}
    \epsilon\ut{f}(t) = \frac{E\ut{tot}(t) - E\ut{tot}(0)}{E\ut{ins}(t)} \ .
    \label{eq:epsf}
\end{equation}
The bottom panel show the feedback efficiency from the same models as the top panel. Initially, after $t\ut{SF}$ when the feedback is activated, $\epsilon\ut{f}(t)$ stays low ($\sim 10^{-6}$) until the expanding bubble is formed at $t \simeq3.3$\,Myr. Then, $\epsilon\ut{f}(t)$ quickly grows to almost $3\times10^{-4}$, and at the end it drops due to the bubble leaving the computational domain. Finally, we define the overall feedback efficiency (or simply the feedback efficiency) $\epsilon\ut{f}$ as a maximum over the whole time evolution of a given model.

\begin{figure}
\includegraphics[width=\columnwidth]{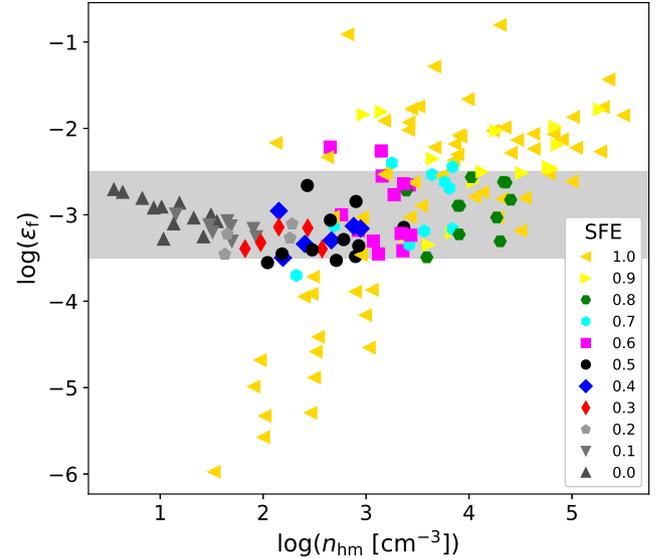}
\caption{\label{fig:efrac} Logarithm of the feedback efficiency $\epsilon\ut{f}$ as a function of $n\ut{hm}$. The shaded area indicates $\epsilon\ut{f}$ values of $10^{-3.5}-10^{-2.5}$.}

\end{figure}

Fig.~\ref{fig:efrac} shows $\epsilon\ut{f}$ as function of $n\ut{hm}$ for all setup~B models. The color and shape of symbols represent the SFE of a given model. Interestingly, if we disregard models with SFE close to unity, majority of models have $\epsilon\ut{f}$ within $10^{-3.5}-10^{-2.5}$. There is a weak trend with $n\ut{hm}$, making $\epsilon\ut{f}$ higher for the high $n\ut{hm}$ (as well as for the higher SFE) with $\epsilon\ut{f}$ reaching $10^{-2}$ for the highest $n\ut{hm} \sim 10^4 - 10^5$\,cm$^{-3}$. Additionally, models with the lowest $n\ut{hm} \lesssim 10$\,cm$^{-3}$ seem to have slightly higher values of $\epsilon\ut{f}$, too.

\section{Discussion}
\label{discuss}


The presented model includes many limitations making its direct comparison to observations difficult; it should be understood rather as a toy model exhibiting some interesting properties improving our insight into how the feedback regulates star formation. Most of the limitations come from the imposed spherical symmetry. However, this also makes the model simple and hence computationally cheap allowing calculations with many parameter combinations, and relatively easy to interpret. Moreover, the model includes the (probably) three most important forms of pre-SN stellar feedback, i.e. the ionising radiation, stellar winds and the radiation pressure, implemented on a relatively detailed level. The obtained results can also be used as a sub-grid model for simulations of star formation and galaxies on larger scales \citep[e.g.][]{2015MNRAS.454..238W}.

One of the most important processes that cannot be properly modelled in 1D is the turbulence in molecular clouds, and it is known to be an important agent regulating the star formation \citep[e.g.][]{2004ARA&A..42..211E,2004RvMP...76..125M}. Its impact on the star formation is both positive and negative. On one hand, it provides an additional pressure and it prevents a monolithic collapse. On the other hand, the turbulence can also lead to density enhancements within the cloud, triggering the local collapse \citep{2000ApJ...535..887K, 2012ApJ...761..156F}. Numerical simulations collectively predict an anticorrelation between SFE and the virial parameter ($\avir$), the latter being function of the (initial) turbulent velocity that defines the level of the cloud boundedness \citep{2004MNRAS.347L..36C, 2012ApJ...759L..27P, 2015MNRAS.451.3679B, 2016ApJ...829..130R, 2016MNRAS.461.2953H, 2021ApJ...911..128K}. From the observational point of view, however, solid conclusions on the negative effect of turbulence on SFE can not be drawn. For example, $\avir$ shows no correlation with SFE in Galactic molecular clouds \citep{2016ApJ...831...73V} nor it is linked to the likelihood of clouds in the LMC hosting young stellar objects \citep{2011ApJS..197...16W}.

More generally, the assumption of spherical symmetry does not allow to include the cloud density structure. It can impact the efficiency of the feedback in dispersing the gas and hence alter the SFE. A highly-structured medium with cavities and low-density channels can allow a significant fraction of photons and momentum to escape reducing thus the coupling of the feedback forces to the gas \citep{2014MNRAS.442..694D,2013MNRAS.431.1337R,2017ApJ...850..112R,2022MNRAS.509..272C}. 

In addition, a shell that undergoes fragmentation may trigger star formation in sites other than the centre \citep[][]{2003A&A...411..397T}, thereby altering the SFE of the cloud. Processes that can attenuate the impact from the stellar winds, such as mixing and cooling in the interface between the shocked gas regime and the ionized gas \citep[e.g.][]{2003ApJ...590..306D,2021ApJ...914...89L} should be further taken into consideration.

Finally, the role of the environment in regulating the SFEs should be taken into account, as it governs the systematic differences between GMCs in different galaxies \citep{2013ApJ...779...46H} and at different galactocentric radii \citep{2012MNRAS.425..720S,2014ApJ...784....3C,2020ApJ...901L...8S}. Coupled to the galactic potential and large-scale dynamics such as the shear, GMCs experience changes in their morphological and kinematic state \citep{2009ApJ...700L.132K,2019MNRAS.484.5734K} that ultimately impact their lifetimes and SFEs \citep{2018MNRAS.476.3688J,2018MNRAS.475.1791C, 2020ApJ...892...73M}. Other external mechanisms that regulate the cloud-scale properties and star formation include compressive shocks induced by nearby star-forming activity \citep{2012ApJ...750..136W} and the cloud-cloud collisions \citep{2021PASJ...73S...1F, 2011MNRAS.413.2935D}.

\section{Conclusions}
\label{concl}

We performed 1D radiation-hydrodynamic simulations to describe the interplay between the pre-SN feedback and the self-gravity of a molecular cloud. The feedback includes the ionising radiation, stellar winds and the radiation pressure on gas and dust from ionising and non-ionising radiation. We study two setups addressing different objectives. Setup~A runs start with the star cluster existing from the beginning, with a fixed mass given by parameter SFE$_0$ and the cloud has always uniform density. The aim is to revisit the semi-analytic work by R18 with a model including the cloud self-gravity and dynamics. In setup~B runs, the star cluster forms from the gas entering the central region, and this allows to study SFE as a function of the model parameters. The cloud density is higher in the centre with various slopes, because a uniform cloud stays uniform during the collapse and does not result in astrophysically meaningful results (leads always to SFE $=1$). Our results are synopsized as follows.


\begin{enumerate}
  \item Comparison to R18 studying a similar model shows that a star cluster with $3-4$ times higher mass is needed to disrupt the uniform cloud by the feedback. Alternatively formulated, for the same star cluster, the cloud needs to have $\sim 5$ times lower density to get disrupted in our model than in R18. This difference is almost entirely due to the self-gravity causing the collapse of the cloud not overrun yet by the shell, which is included in our model while it is not in R18.
  

  \item Calculations with higher metallicity exhibit more efficient feedback. In our setup~A runs with solar metallicity, the feedback bubble expands slightly faster and the cloud disruption is easier than in runs with LMC metallicity ($0.4$\,Z$_\odot$). It is due to the higher mechanical luminosity of stellar winds of stars with higher metallicity.


  \item Our model is qualitatively consistent with the suggestion of R18 that the R136 star cluster in NGC\;2070 might have been formed by the re-collapse of the shell. Assuming the mass of the 30~Doradus natal cloud $M\ut{cld}=10^{6.5}$ M$_{\odot}$, the re-collapse scenario may take place with cloud density $n_0\lesssim1000$\,cm$^{-3}$.

  \item The star formation efficiency is a function of a single parameter combining the all properties of the cloud in our model. This parameter is the mean density within the cloud half-mass radius, $n\ut{hm}$, and it is given by the cloud mass, radius and the slope of its density profile. The star formation efficiency in our setup~B runs, i.e. the mass of the formed star cluster normalized by the initial mass of the cloud, follows approximately the relation $\log(\mathrm{SFE}) = -5.43(n\ut{hm}/\mathrm{cm}^{-3})^{-0.46}$.
  


  \item The feedback efficiency, i.e. a fraction of the energy retained by gas from the total energy inserted by the feedback, is estimated to be approximately $10^{-3}$, with almost flat distribution according to $n\ut{hm}$.
  
\end{enumerate}

Future transition of our model to the three dimensions will enable the inclusion of turbulence, address non-spherical instabilities, and allow star formation to take place in sites other than the centre e.g.\ in the compressed shell, towards an thorough insight into the processes that govern the internal properties of GMCs.

\section*{Acknowledgments}%

M.K., R.W., J.P., and S.E. acknowledge financial support from the Czech Science foundation GA\v{C}R under grant number 19-15008S and by the Astronomical Institute of the Czech Academy of Sciences through
the project RVO:67985815. M.K. has also received funding from the European Union's 
Framework Programme for Research and Innovation Horizon 2020 (2014-2020) under the Marie Sk\l{}odowska-Curie Grant Agreement No. 823734. S.M.-G. and G.T.-T. were partly supported by CONACYT-M\'exico research grant A1-S-28458. S.M.-G. also acknowledges support by CONACYT through project n.482 of the “Programa Investigadoras e Investigadores por M\'exico”.

\section*{Data availability}

The data underlying this article will be shared on reasonable request to the corresponding author.

\bibliographystyle{mnras}
\bibliography{kourniotis22}
\clearpage

\appendix

\section{Dependence on resolution}
\label{app:resolution}

\begin{figure}
\centering
\includegraphics[width=\columnwidth]{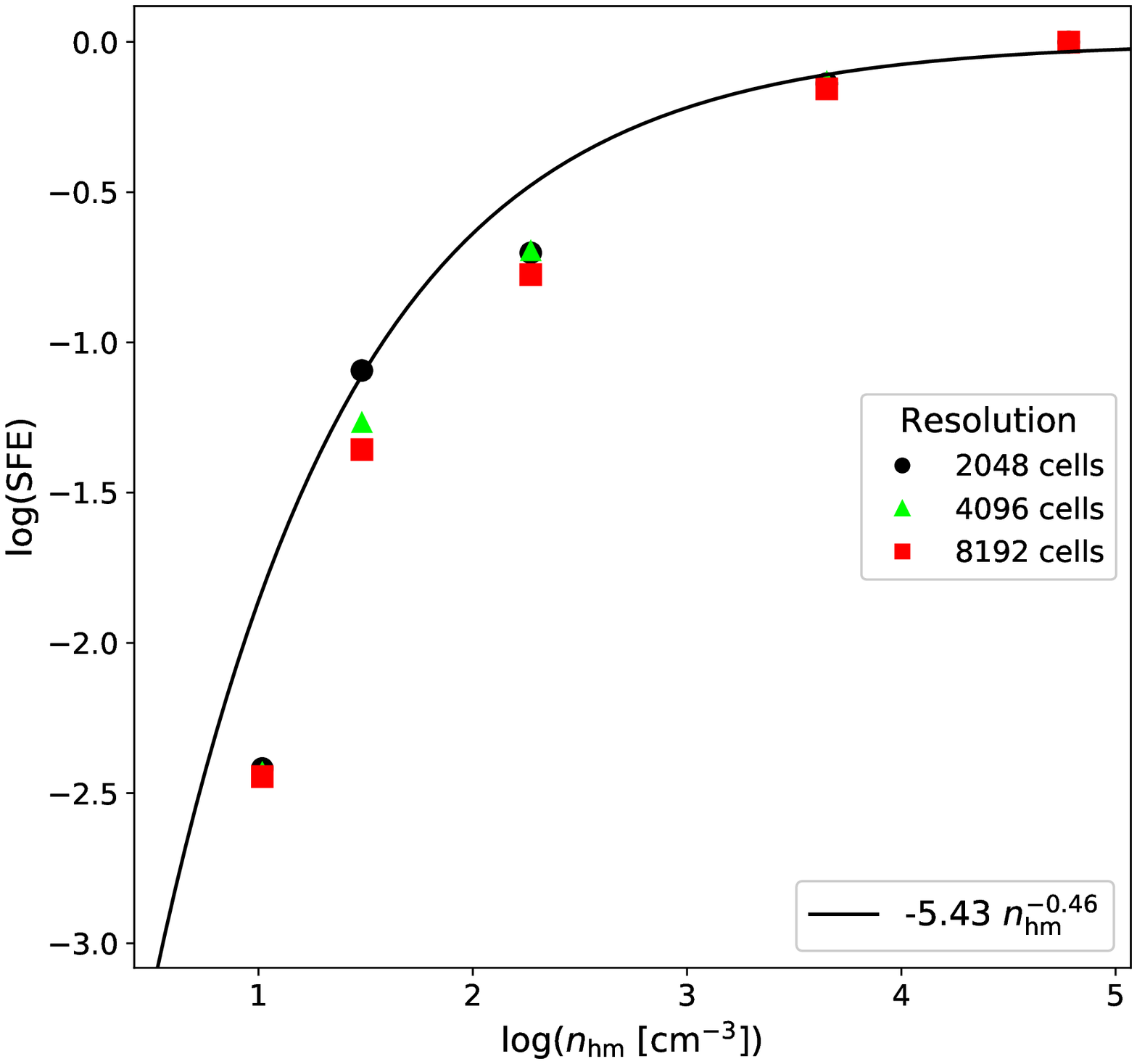}
\caption{Same as in Fig. \ref{sfe_rho}, for five cloud models under setup~B, which were tested under three refinement levels (see text). For reference, we display the fit curve to the entire set of setup~B models (Eq. \ref{eq:sfe_rho}).}
\label{resol_app}
\end{figure}

We explore the effect that the grid resolution has on our results. We run setup~B simulations with five sets of input parameters for the cloud mass, radius, and density slope; ($M\ut{cld}$, $R\ut{cld}$, $\beta$) = (10$^{4.5}$ \MSun, 20 pc, 1.4),  (10$^{4.5}$ \MSun, 30 pc, 0.8), (10$^{4.5}$ \MSun, 20 pc, 1.2), (10$^{5.5}$ \MSun, 20 pc, 1.6), and (10$^{6}$ \MSun, 10 pc, 1.8). For our test purposes, the computational domain was divided into 2048, 4096, and 8192 grid cells. Following our description in \S\ref{sec:setupB}, we calculated the half-mass densities of the model clouds and display them against the resulting SFEs in Fig. \ref{resol_app}, similar as in Fig. \ref{sfe_rho}. For a given cloud model, the discrepancy in the SFEs due to the different resolutions is shown to be, in general, within $\sim 0.1$ dex.

\begin{landscape}

\section{Setup~B runs - Entire set}
\label{app:full_param}

Calculations over the entire set of cloud parameters are displayed in Fig. \ref{sfr_app}, showing the dependence of SFE on $M\ut{cld}$ and $R\ut{cld}$. Generally, SFE is lower for lower $M\ut{cld}$ and higher $R\ut{cld}$. These relations are valid both when comparing models with other parameters fixed, and also for $\mathrm{SFE}(\beta\ut{min})$. This is expected, because more strongly gravitationally bound clouds need more feedback energy and hence higher mass star cluster to be disrupted. Additionally, $\beta\ut{min}$ gets higher for higher $M\ut{cld}$ and lower $R\ut{cld}$. Note that, the lowest measured SFE in models with $M\ut{cld} = 10^{4.5} - 10^{5.0}$\,\MSun and $R\ut{cld}=30-40$\,pc is constrained by the threshold mass of the cluster for initiating feedback, $M\ut{SF}$ (dashed line).

\begin{figure}
\centering
\includegraphics[width=\linewidth]{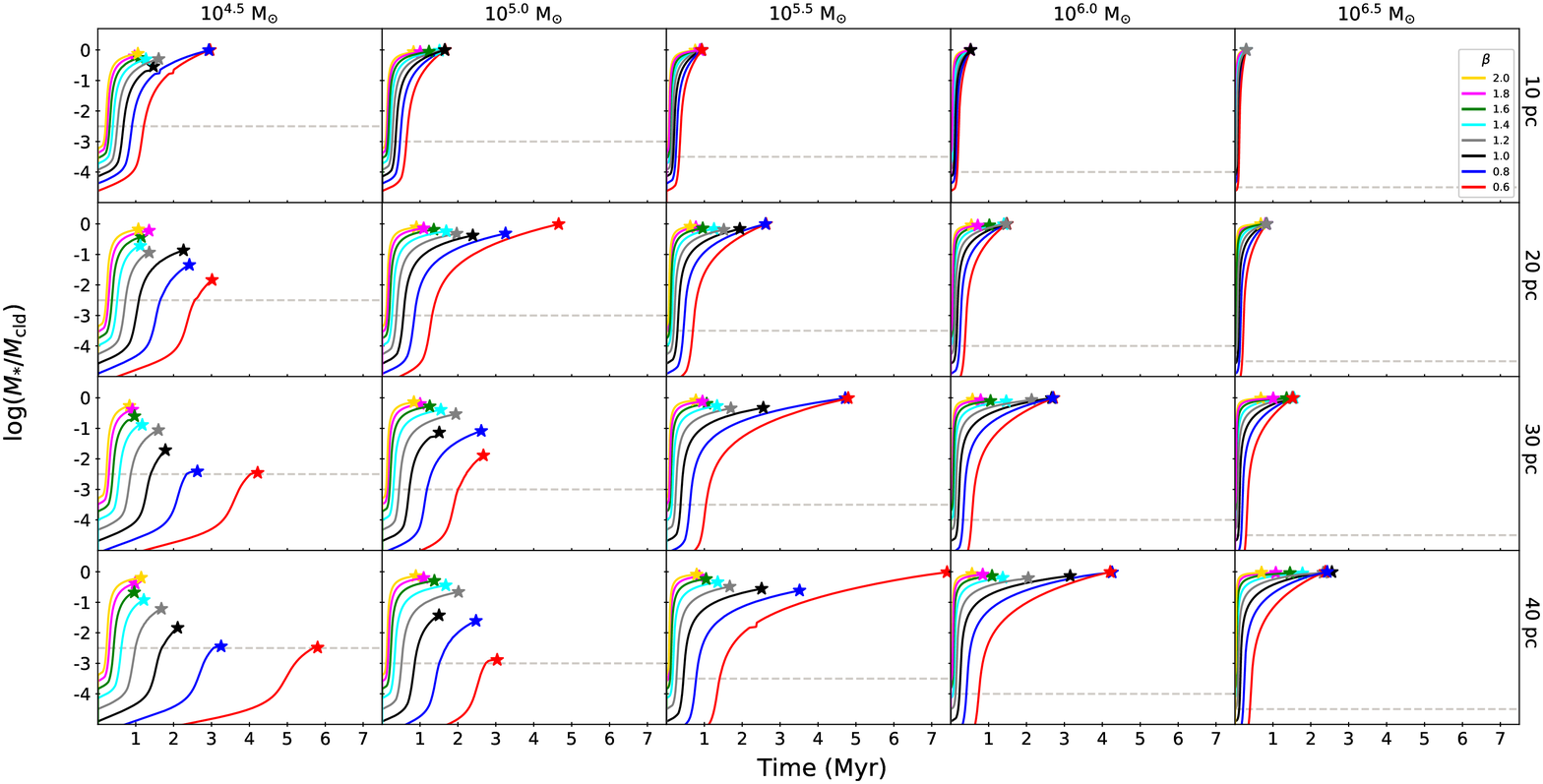}
\caption{Same as in Fig. \ref{sf_rate}, for the different cloud models of the explored parameter space. The calculations on the growth of the stellar mass are color coded with respect to the steepness $\beta$ of the density profile, with designations given in the upper right panel.}
\label{sfr_app}
\end{figure}

\label{lastpage}

\end{landscape}

\end{document}